\pdfoutput=1
\documentclass{aastex701}

\usepackage{xcolor}
\newcommand{\rev}[1]{#1}

\begin{document}

\title{Solar Orbiter SEP Dropout during a Magnetic Cloud with Evidence for Strong Connectivity Gradients}

\author[orcid=0000-0001-7381-6949, gname=Radoslav, sname=Bucik]{Radoslav Bu\v{c}\'{i}k}
\affiliation{Southwest Research Institute, San Antonio, TX 78238, USA}
\email[show]{radoslav.bucik@swri.org}  

\author[orcid=0000-0002-5705-9236,gname=Raul, sname=Gomez-Herrero]{Ra\'{u}l G\'{o}mez-Herrero} 
\affiliation{Universidad de Alcal\'{a}, Space Research Group, 28805 Alcal\'{a} de Henares, Spain}
\email{raul.gomezh@uah.es}

\author[orcid=0000-0003-0508-4912,gname=Samuel, sname=Hart]{Samuel T. Hart} 
\affiliation{Southwest Research Institute, San Antonio, TX 78238, USA}
\email{samuel.hart@swri.org}

\author[orcid=0000-0001-9323-1200,gname=Maher, sname=Dayeh]{Maher A. Dayeh}
\affiliation{Southwest Research Institute, San Antonio, TX 78238, USA}
\affiliation{University of Texas at San Antonio, San Antonio, TX 78249, USA}
\email{maher.aldayeh@swri.org}

\author[orcid=0000-0001-6119-0221,gname=Nariaki, sname=Nitta]{Nariaki V. Nitta}
\affiliation{Lockheed Martin Advanced Technology Center, Palo Alto, CA 94304, USA}
\email{nitta@lmsal.com}

\setcounter{footnote}{0}


\begin{abstract}

We analyze an impulsive solar energetic particle (SEP) event observed by Solar Orbiter at 0.93\,au on 2022 December 24 that exhibits a pronounced intensity dropout between $\sim$07:15 and 10:00\,UT. Pitch-angle distributions show a near-field-aligned beam before the dropout, which disappears abruptly at the dropout onset. At the same time, a weak 100--200\,keV component appears at pitch angles of $\sim$90$^\circ$--180$^\circ$; no comparable enhancement over this pitch-angle range is present at MeV energies. The dropout onset is not accompanied by an abrupt change in the in situ magnetic field or solar wind plasma. Solar imaging associates the SEP event with a jet from a compact source. Ballistic back-mapping, together with Potential Field Source Surface extrapolations and quasi-separatrix-layer proxy diagnostics, indicates that Solar Orbiter nominal\rev{ly} connect\rev{s} \rev{to} the source region, but lies close to strong connectivity gradients where small displacement \rev{may} shift the connection to neighboring open field \rev{lines}. The \rev{in situ} measurements show that the event occurs during a magnetic-cloud passage and that additional dropouts occur later in the event. These results favor an interpretation in which the dropout reflects rapid changes in particle access between adjacent flux tubes, while the magnetic cloud may help preserve the sharp SEP intensity gradients between them.
\end{abstract}



\section{Introduction}\label{sec:int}

Impulsive (${}^3\mathrm{He}$-rich) solar energetic particle (SEP) events often exhibit intensity dropouts, in which ion fluxes decrease rapidly and simultaneously over a broad energy range \citep{2000ApJ...532L..79M,2004ApJ...614..412G}. Dropouts may recur multiple times during a single event and are commonly reported below $\sim 1\,\mathrm{MeV\,nucleon^{-1}}$ \citep{2008ApJ...688.1368C}. In contrast, clear dropouts are less commonly reported in gradual SEP events, consistent with their broader injection regions, although dropout-like signatures have also been discussed in some gradual events, particularly during decay phases \citep[e.g.,][]{2023ApJ...954...26T}. 

Two explanations have been proposed. In one, dropouts result from changes in magnetic connectivity, such that neighboring flux tubes at 1\,au connect to widely separated coronal footpoints and alternately sample SEP-filled and SEP-poor regions near the Sun \citep{2000ApJ...532L..79M,2000ApJ...532L..75G}. In the other, turbulence in the solar wind \rev{can} produce filamentary SEP \rev{intensity patterns} and intermittent transport, generat\rev{ing} dropout-like intensity variations \citep{2003ApJ...597L.169R,2010ApJ...711..980S}. The relative importance of these effects remains debated, and distinguishing observationally between connectivity-driven and turbulence-driven scenarios has proven difficult \citep{2006ApJ...641L..61G,2014ApJ...780...16G}.

\rev{At 1\,au,} dropouts in impulsive SEP events often show no clear correspondence with abrupt changes in the local interplanetary magnetic field (IMF) or solar wind plasma parameters \citep{2000ApJ...532L..79M,2008ApJ...688.1368C}, although associations with stream interfaces, heliospheric current-sheet crossings, or transient structures have been reported in some events \citep{1992ApJ...387..715R,2004ApJ...614..412G,2013ApJ...770...11T}. Solar Orbiter has also observed dropout signatures in impulsive events at different heliocentric distances, including repeated dropouts across multiple injections at $\sim$0.95\,au \citep{2022FrASS...9.9799H,2023FrASS..1048467N} and modest dropouts within an interplanetary coronal mass ejection (ICME)/flux-rope interval at 0.43\,au \citep{2023A&A...678A..98W}.

In this paper we analyze a ${}^3\mathrm{He}$-rich SEP event observed by Solar Orbiter at 0.93\,au on 2022 December 24 that shows a clear dropout. We combine energetic-ion pitch-angle distributions (PADs) with in situ IMF and solar wind plasma measurements, suprathermal electron observations, and solar imaging of the source region. We also evaluate magnetic connectivity using ballistic back-mapping together with Potential Field Source Surface (PFSS) modeling and quasi-separatrix layer (QSL)-proxy diagnostics. This combined approach provides new constraints on the origin of the dropout.

\section{Instruments}\label{sec:ins}

Ion PADs were measured with the Electron Proton Telescope (EPT), and ion composition was examined with the Suprathermal Ion Spectrograph (SIS), both part of the Energetic Particle Detector (EPD) suite \citep{2020AA...642A...7R} on Solar Orbiter \citep{2020AA...642A...1M}. EPT has four viewing directions: one unit points sunward and anti-sunward along the average Parker spiral (EPT-Sun and EPT-ASun; EPT-Sun is oriented 35$^\circ$ west of the spacecraft-Sun line), while the other unit points northward and southward out of the ecliptic (EPT-North and EPT-South). EPT covers $\sim$0.05--6\,MeV. The instrument provides ion fluxes at 1 and 5\,s cadences in 64 logarithmically spaced energy channels. These ion channels are calibrated using the hydrogen response but are not mass-resolved. We therefore refer to them as EPT ion intensities. SIS is a time-of-flight mass spectrometer that measures H through ultra-heavy ions over $\sim$0.1--10\,MeV\,nucleon$^{-1}$. In this study we use measurements from the sunward SIS-a telescope, which points 30$^\circ$ west of the spacecraft-Sun line. Solar wind plasma measurements were obtained from the Solar Wind Analyzer \citep[SWA;][]{2020AA...642A..16O}, specifically from the Proton and Alpha Sensor (PAS) and Electron Analyser System (EAS). Magnetic field measurements were provided by the magnetometer \citep[MAG;][]{2020AA...642A...9H} on Solar Orbiter. The solar source was examined using observations from the Atmospheric Imaging Assembly \citep[AIA;][]{2012SoPh..275...17L} and Helioseismic and Magnetic Imager \citep[HMI;][]{2012SoPh..275..207S} onboard the Solar Dynamics Observatory (SDO). For this event, Solar Orbiter/EUI \citep{2020AA...642A...8R} was in occulter mode.

\section{SEP dropout}\label{sec:sep}

\begin{figure}
\centering
\epsscale{1.15}
\plotone{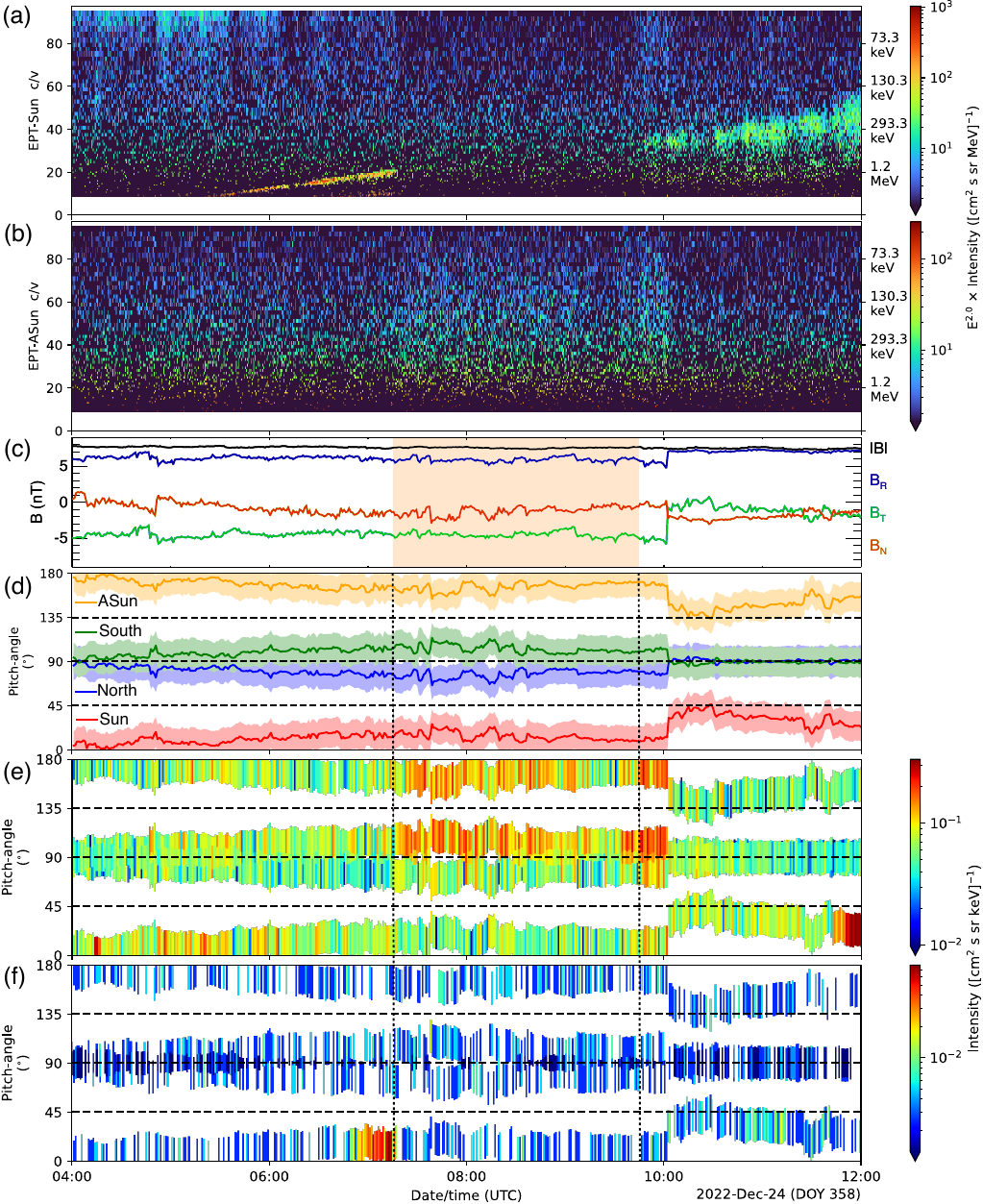}
\caption{(a) EPT-Sun and (b) EPT-ASun inverse-velocity ($c/v$) versus time spectrograms \rev{of $E^{2}J(E, t)$ } on 2022 December 24. (c) IMF (1 minute) magnitude $|B|$ and components in RTN coordinates. Shading denotes dropout period. (d) Pitch-angle coverage of \rev{the four} EPT \rev{telescopes}. (e) Pitch-angle distribution of EPT ions in 103.36--203.33\,keV (60 s averaged). (f) Pitch-angle distribution of EPT ions in 1.03--1.46\,MeV (60 s averaged). Two black vertical dotted lines mark the dropout period.}
\label{fig:pa}
\end{figure}

Figure~\ref{fig:pa} \rev{provides an overview of} the EPT \rev{and} MAG measurements \rev{during the SEP event}. Figure~\ref{fig:pa}(a--b) shows 30\,s averaged $\sim$52\,keV to 6.4\,MeV ion-channel intensities $J(E, t)$ in an inverse-velocity versus time spectrogram for the two EPT telescopes (Sun, ASun) on 2022 December 24 between 04--12\,UT. A clear dispersive SEP onset is observed in the Sun telescope over $\sim$05--07\,UT, with the highest-energy ions arriving first and lower-energy ions later. The other three telescopes remain near background prior to the dropout, indicating a strongly directional beam confined to the Sun telescope’s field of view. A sharp intensity dropout occurs near $\sim$07:15\,UT, followed by a more gradual recovery of ions at $\sim$09:45--10:00\,UT restricted to the Sun telescope. A faint enhancement is present in ASun during the dropout. South shows a similar \rev{faint enhancement}, while North lacks such a feature (Fig.~\ref{fig:pa}e). This event is ${}^3\mathrm{He}$-rich \rev{as indicates He mass spectrogram in Fig.~\ref{fig:mu}c}.



Figure~\ref{fig:pa}c shows the IMF components in RTN coordinates. $|B|$ remains nearly steady and $B_\mathrm{R} > 0$ (outward polarity) throughout the interval, with a brief directional change dominated by $B_\mathrm{T}$ near the end of the dropout. Figure~\ref{fig:pa}d shows the pitch-angle coverage of the four EPT telescopes, computed from the IMF in the spacecraft frame. Prior to and throughout the dropout, the Sun telescope samples small pitch angles, $\alpha\sim$0$^\circ$--30$^\circ$, whereas ASun samples $\alpha\gtrsim135^\circ$, and North and South primarily sample $\alpha\sim90^\circ$. After the dropout, the Sun-telescope coverage shifts modestly to $\alpha\sim$20$^\circ$--50$^\circ$, consistent with a change in IMF direction. 

Figure~\ref{fig:pa}(e--f) shows 60\,s averaged PADs for EPT ions over 103--203\,keV and 1.03--1.46\,MeV. Each interval is formed by summing adjacent EPT energy channels. Prior to the dropout, the 1.03--1.46\,MeV enhancement is confined to small $\alpha$ in the Sun telescope, demonstrating a strongly field-aligned beam with no counter-streaming. During the dropout, the near-field-aligned population abruptly diminishes, while a weak intensity increase appears at 103--203\,keV over $\alpha\sim$90$^\circ$--180$^\circ$ (ASun and South). No such enhancement over this pitch-angle range is observed at 1.03--1.46\,MeV. Because the EPT pitch-angle coverage remains essentially unchanged across the dropout onset (Figure~\ref{fig:pa}d), this behavior is unlikely to be produced by changing viewing geometry. No abrupt change in IMF direction is seen until near the end of the dropout.

\section{Solar source}\label{sec:sol}

The event is associated with type III radio \rev{emission} near 04:10\,UT and a solar energetic electron event observed by EPT-Sun. The type III \rev{emission was} identified in browse radio spectrograms from Solar Orbiter/RPW \citep{2020AA...642A..12M}, STEREO-A/S/WAVES \citep{2008SSRv..136..487B}, and Wind/Waves \citep{1995SSRv...71..231B}. The NOAA Space Weather Prediction Center (SWPC) Edited Events list reports a GOES C4.0 flare (start 04:08 UT; peak 04:14\,UT) from AR 13169. The in-situ dispersive SEP onset at Solar Orbiter is consistent with interplanetary travel time following the $\sim$04:10\,UT solar activity.

\begin{figure}
\centering
\epsscale{0.80}
\plotone{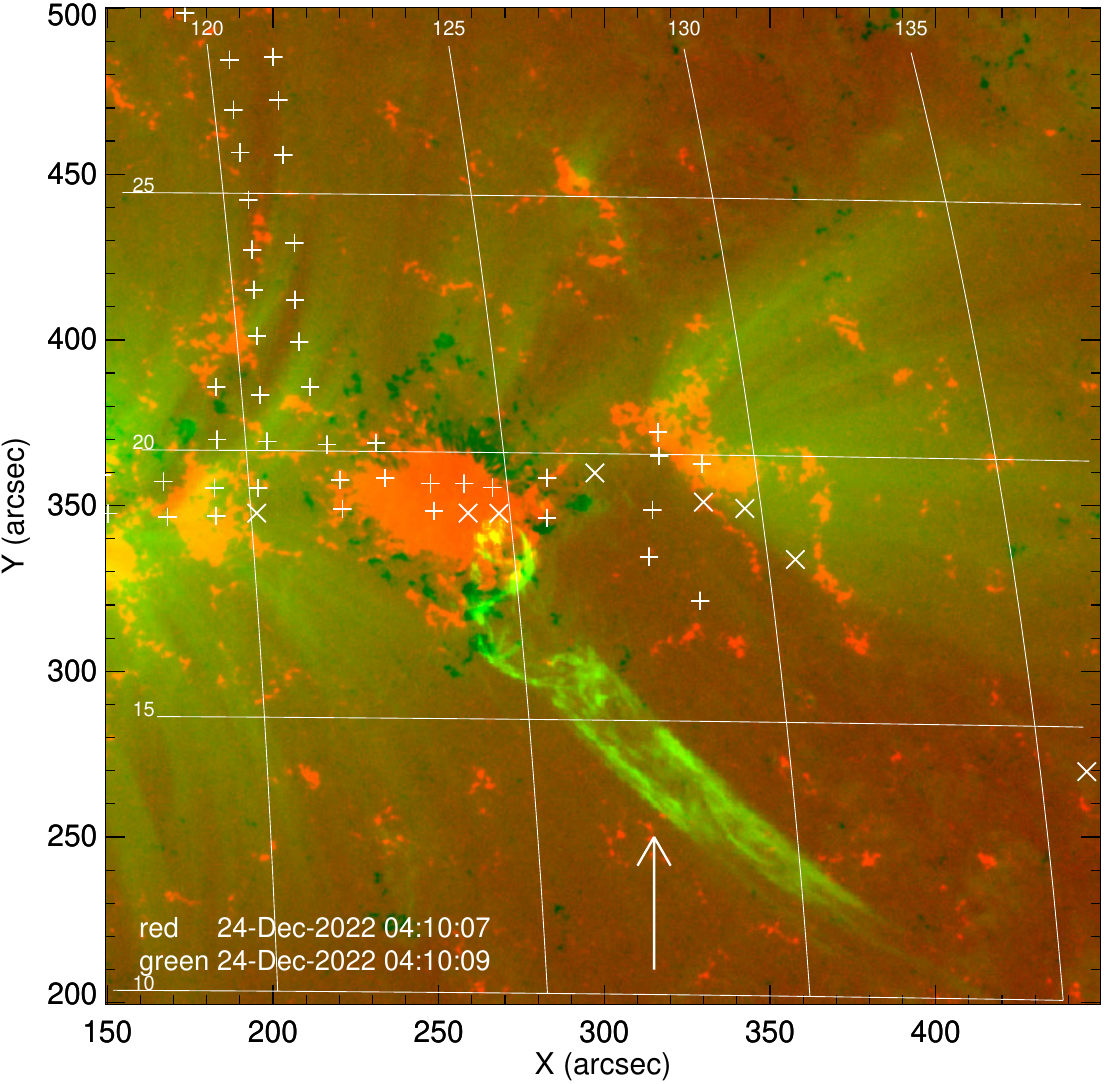}
\caption{SDO image of the solar source at time of the event-associated type III radio burst. AIA 171\,{\AA} is shown in green, and the HMI line-of-sight magnetic field (scaled to $\pm$150\,G) in black and red tinted colors. Jet is marked by arrow; a 5$^\circ$ Carrington grid is overlaid. \rev{The $\times$ symbols mark photospheric footpoints of field lines open to the ecliptic, and the $+$ symbols mark footpoints of field lines open to other latitudes at the source surface (Section~\ref{sec:mag}).}}
\label{fig:ss}
\end{figure}

The solar source is a jet with a base centered at Stonyhurst longitude 16.6$^\circ$ (Carrington longitude 124.6$^\circ$) and latitude $\sim$17.6$^\circ$, as observed in AIA 1700\,{\AA} (upper photosphere/low chromosphere) and AIA 94\,{\AA} (hot coronal emission) at 04:10\,UT. AIA 1700\,{\AA} reveals two compact, co-temporal brightenings at the jet base, while AIA 94\,{\AA} shows a single broader brightening (longitude extent $\sim$1.3$^\circ$; latitude extent $\sim$2.9$^\circ$) overlapping the 1700\,{\AA} brightenings. The jet is located at the edge of a large sunspot in AR 13169. Figure~\ref{fig:ss} displays the source in composite images of AIA 171\,{\AA} ($\sim$1\,MK coronal plasma) and the HMI line-of-sight magnetogram. Such jet activity and type III radio emission are commonly associated with ${}^3\mathrm{He}$-rich SEP events \citep[e.g.;][]{2006ApJ...650..438N,2020SSRv..216...24B}.

\section{Magnetic connections}\label{sec:mag}

\subsection{Solar Orbiter connections} \label{subsec:con}

We examine Solar Orbiter (SO) magnetic connectivity on 2022 December 24 between 03--12\,UT. Over this interval, SO is at $r\sim0.93$\,au, heliographic latitude 4.48$^\circ$--4.52$^\circ$, and Carrington longitude 84.5$^\circ$--89.6$^\circ$. We estimate the source-surface footpoint at $r_\mathrm{SS} = 2.5$\,R$_{\odot}$ using standard Parker-spiral ballistic back-mapping \citep[e.g.,][]{1958ApJ...128..664P,1973SoPh...33..241N}, assuming a constant radial solar wind speed between $r_\mathrm{SS}$ and Solar Orbiter: $\phi_\mathrm{SS}=\phi_\mathrm{SO}+\Omega(r-r_\mathrm{SS})/v_\mathrm{sw}$. Here, $\phi_\mathrm{SO}$ and $\phi_\mathrm{SS}$ are the Carrington longitudes of Solar Orbiter and its source-surface footpoint, respectively, $r$ is the spacecraft heliocentric distance, $\Omega$ is the solar rotation rate, and $v_\mathrm{sw}$ is the 1\,hr averaged solar wind speed. The resulting source-surface footpoint is then traced to the photosphere using the SolarSoft PFSS package \citep{2003SoPh..212..165S}, where synoptic magnetograms are available every 6\,hr. 

We evaluate the effect of a longitudinal uncertainty of $\pm$10$^\circ$ in the ballistic mapping \citep[e.g.,][]{1973SoPh...33..241N} by tracing field lines from source-surface longitudes $\phi_\mathrm{SS}-10^\circ$ (SO$-$10) and $\phi_\mathrm{SS}+10^\circ$ (SO$+$10). For the 03--05\,UT interval (around injection) and the 07--10\,UT interval (during the dropout), we quantify the sensitivity to $v_\mathrm{sw}$ variability by calculating $\phi_\mathrm{SS}$ using the minimum and maximum 1\,hr averaged solar wind speeds in each interval. The corresponding speed ranges, 377--381\,km\,s$^{-1}$ and 359--383\,km\,s$^{-1}$, give source-surface longitude uncertainties of $\sim\pm0.3^\circ$ and $\sim\pm2^\circ$, respectively.

\subsection{Magnetic connectivity gradients  }\label{subsec:gra}

\begin{figure}
\centering
\plotone{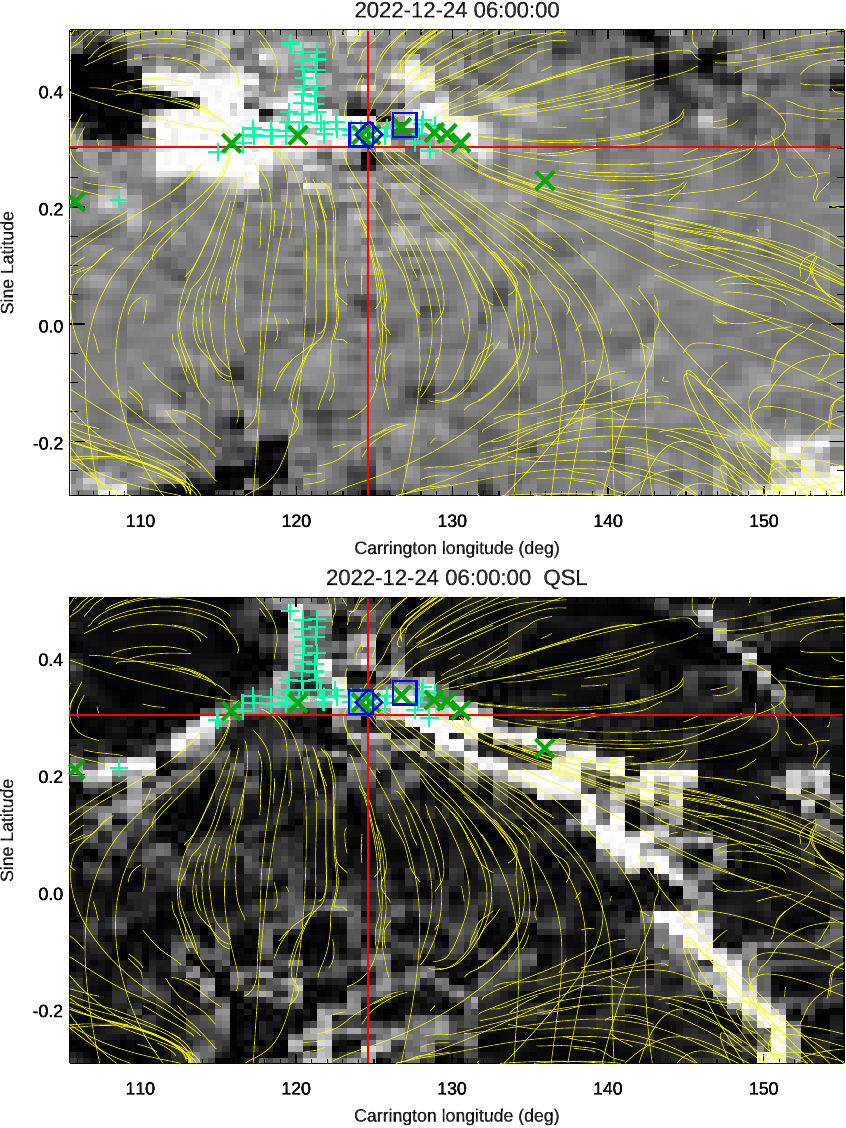}
\caption{PFSS photospheric maps at 2022-12-24 06:00\,UT. Top: The radial magnetic field, scaled to $\pm$30\,G. Bottom: A QSL-proxy map (Section~\ref{subsec:gra})\rev{, where brighter pixels indicate stronger connectivity gradients}. Yellow curves indicate a subset of traced closed field lines. Symbols mark photospheric footpoints of open field lines traced from a uniform photospheric seed grid and classified by the latitude of their source-surface endpoint: open-to-ecliptic ($|\lambda_\mathrm{SS}|\le7^\circ$; \rev{green $\times$ symbols}) and open-to-other source-surface latitudes (\rev{cyan $+$ symbols}). Both \rev{symbol types} denote positive polarity. The blue diamond marks Solar Orbiter (SO) and blue squares mark $\pm$10$^\circ$ Parker-spiral longitude offsets (SO$-$10, SO$+$10). The red crosshair intersects at the EUV jet-base center.}
\label{fig:qs}
\end{figure}

We quantify magnetic-connectivity gradients using a QSL proxy computed from PFSS field-line mappings. Field lines are traced from a uniform grid of photospheric cell centers. For each start location ($\phi$, $\theta$), where $\phi$ and $\theta$ denote heliographic longitude and colatitude, we construct unit vectors at the start point and at the traced endpoint (the conjugate photospheric footpoint for closed field lines or the source-surface intersection for open field lines). We estimated local mapping gradients using forward finite differences between neighboring start cells in the longitudinal (east) and latitudinal (north) directions by comparing great-circle angular separations of the corresponding endpoint vectors.

Denoting by $\Delta x_\mathrm{s}$, $\Delta y_\mathrm{s}$ the angular separations between neighboring start vectors, and by $\Delta x_\mathrm{f}$, $\Delta y_\mathrm{f}$ the separations between the mapped endpoints, we defined 
\begin{equation}
Q_{\mathrm{proxy}} \equiv
\sqrt{\left(\frac{\Delta x_{\mathrm{f}}}{\Delta x_{\mathrm{s}}}\right)^2
+\left(\frac{\Delta y_{\mathrm{f}}}{\Delta y_{\mathrm{s}}}\right)^2}\,
\cos\lambda
\label{eq:qproxy}
\end{equation}
where $\lambda$ is heliographic latitude. This proxy highlights QSL-like ridges where the field-line mapping varies rapidly with position \citep[e.g.,][]{2002JGRA..107.1164T,2007ApJ...660..863T}.

Figure~\ref{fig:qs} shows zoomed views (latitude $-17^\circ$ to 30$^\circ$, longitude 105$^\circ$ to 155$^\circ$) of the radial magnetic field and QSL-proxy maps around the source region at 06:00\,UT (PFSS synoptic magnetogram time 06:04\,UT). The brightest filamentary structures in the QSL-proxy map mark strong connectivity gradients. At this time, SO and SO$-$10 trace to photospheric footpoints within the jet base that contains open-to-ecliptic field lines of the measured IMF polarity, whereas SO$+$10 traces to open-to-ecliptic footpoints displaced from the jet base. Because these footpoints lie near strong connectivity gradients, ballistic-mapping uncertainty and/or coronal-field time dependence can shift the estimated footpoint between neighboring open-field connectivity regions of the same polarity.

\subsection{\rev{Temporal variation of Solar Orbiter photospheric footpoints}}\label{subsec:evo}

\begin{figure}
\centering
\epsscale{0.85}
\plotone{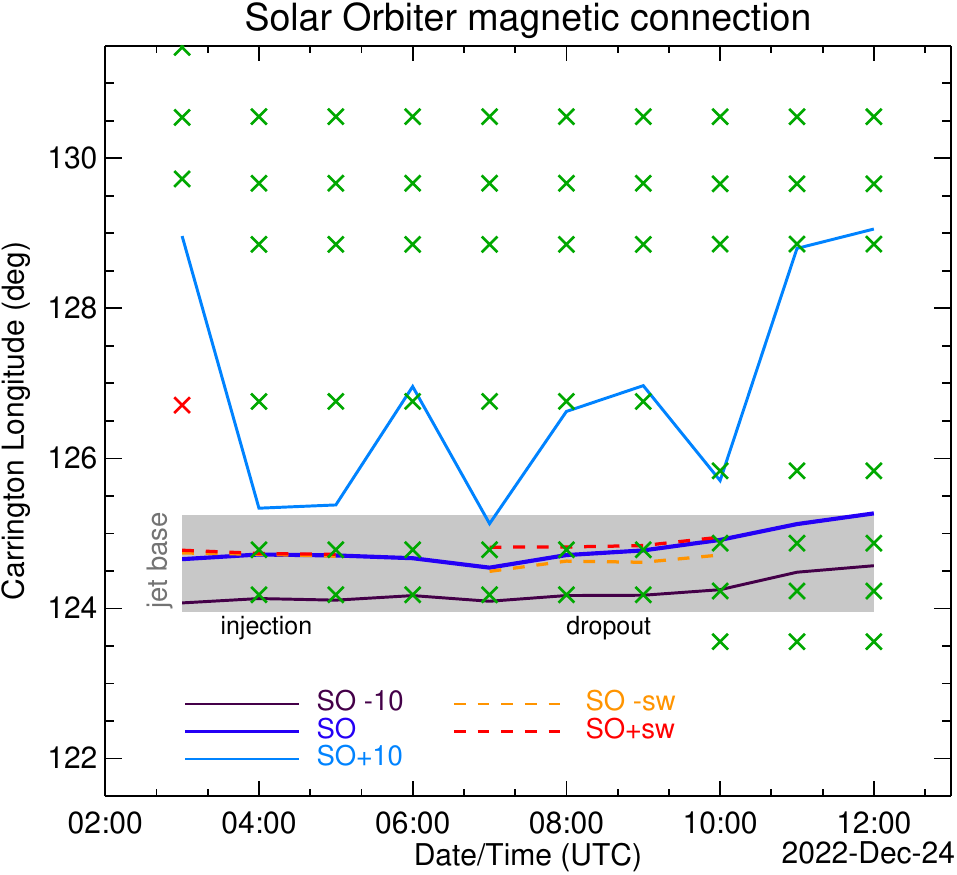}
\caption{Carrington longitude of SO photospheric footpoints versus time. Solid curves show SO, SO$-$10, and SO$+$10. Dashed curves \rev{SO$-$sw and SO$+$sw show} the longitude uncertainty from measured $v_\mathrm{sw}$ variability during the injection (03--05\,UT) and dropout (07--10\,UT) intervals. \rev{The $\times$ symbols} mark photospheric footpoints of open-to-ecliptic field lines ($|\lambda_\mathrm{SS}|\le7^\circ$), colored by polarity \rev{(green for positive, red for negative)}. Photospheric footpoints whose source-surface endpoints lie at other latitudes (shown in Fig.~\ref{fig:qs}) are omitted for clarity. The gray band denotes the EUV jet-base longitude range.}
\label{fig:ev}
\end{figure}

Figure~\ref{fig:ev} shows the \rev{temporal variation} of the SO photospheric footpoint on 2022 December 24 between 03--12\,UT, plotted over a restricted longitude interval (121.5$^\circ$--131.5$^\circ$) around the jet source. \rev{For each time, the PFSS tracing uses the closest available 6 hr field: 00:04\,UT at 03\,UT, 06:04\,UT from 04--09\,UT, and 12:04\,UT from 10--12\,UT.} Over this period, SO and SO$-$10 remain closely spaced, and their traced footpoints stay within the jet-base longitude range, which from $\sim$04\,UT contains open field lines whose source-surface endpoints lie within the ecliptic band ($|\lambda_\mathrm{SS}|\le7^\circ$). In contrast, SO$+$10 exhibits large variability because its photospheric footpoint lies near regions of strong connectivity gradients. As a result, it frequently shifts to open field lines outside the jet base whose source-surface endpoints lie within the ecliptic band, including during the dropout interval. The longitude uncertainty associated with the observed solar wind speed variability is small, so the SO$-$sw, SO, and SO$+$sw are nearly co-located in Fig.~\ref{fig:ev}. 

Within the longitude window of Fig.~\ref{fig:ev}, 74\% of open-to-ecliptic footpoints fall within the EUV jet-base latitude range (16.15$^\circ$--19.05$^\circ$). The remainder lie only slightly outside ($\lesssim0.6^\circ$), comparable to the PFSS latitude discretization ($\Delta\lambda\sim0.93^\circ$ per pixel; half-pixel $\sim0.47^\circ$), and are therefore consistent with the source latitude.

\section{Solar wind during the SEP dropout}\label{sec:sow}

\begin{figure}
\centering
\epsscale{1.16}
\plotone{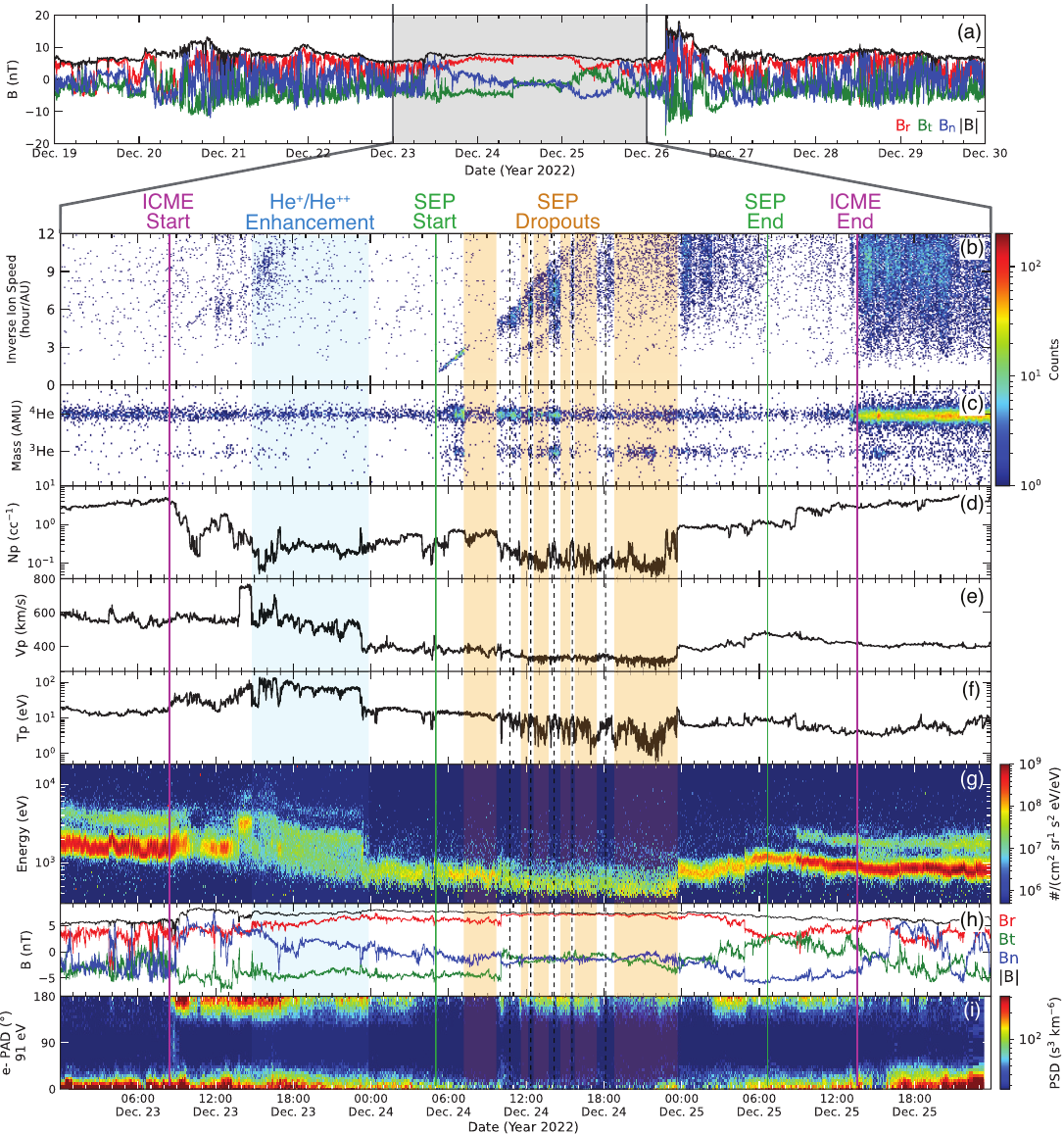}
\caption{Solar wind and IMF conditions around the 2022 December 24 ${}^3\mathrm{He}$-rich SEP event. (a) Local IMF for 11 days surrounding the event. The shaded interval marks the ICME passage. (b--i) Three days of measurements within the ICME. (b) SIS C--Fe \rev{inverse ion-speed spectrogram}. The SEP event start and stop are marked by the vertical green lines, and the SEP dropouts are indicated by the orange shaded regions. \rev{(c) SIS He mass spectrogram at 0.4--10\,MeV\,nucleon$^{-1}$.} (d--f) SWA-PAS solar wind proton moments. The large variability \rev{during the SEP event} likely \rev{reflects} the \rev{the very} low \rev{proton} densities. (g) Solar wind energy spectrogram. Possible He$^{+}$ enhancement is indicated by the blue shaded region. (h) IMF components and magnitude. (i) SWA-EAS  91\,eV suprathermal electron pitch angle distributions. Dashed black lines mark intervals of electron depletion during the later dropout sequence.}
\label{fig:mu}
\end{figure}

Section~\ref{sec:sep} shows that the SEP dropout onset is not caused by a sudden change in EPT pitch-angle coverage, because the IMF does not rotate abruptly (Fig.~\ref{fig:pa}c--d). Here we examine the event using a broader set of in situ measurements. Figure~\ref{fig:mu} shows the IMF over $\pm5$ days around the event (Fig.~\ref{fig:mu}a) and zooms in on a three-day interval spanning the SEP event (Fig.~\ref{fig:mu}b--i). Roughly one day before the event, the IMF fluctuations decrease markedly, consistent with the passage of an ICME/magnetic-cloud interval \citep[e.g.,][]{2006SSRv..123...31Z} that encompasses the investigated SEP event.

Figure~\ref{fig:mu}b shows an inverse-ion-speed spectrogram of SIS heavy ions (C--Fe). The dropout interval studied in detail above is the first and deepest of several $\sim$1-hour SEP dropouts during this event. Later dropouts are generally less pronounced (orange shaded regions). \rev{Figure~\ref{fig:mu}c shows a He mass spectrogram. A ${}^3\mathrm{He}$ enhancement is present during the final dropout interval, corresponding to dispersive injections dominated by ${}^3\mathrm{He}$ (not shown).} Figures~\ref{fig:mu}(d--f) show 1-minute solar wind proton moments derived from the 4-second energy flux distributions in Fig.~\ref{fig:mu}g. At the ICME onset (first vertical purple line), the proton density drops sharply from $\sim$5\,cm$^{-3}$ to $\sim$0.1\,cm$^{-3}$ and remains below $\sim$1\,cm$^{-3}$ for the next two days. After the density decrease, the solar wind energy spectrum shows an additional peak (blue shading) near twice the $E/q$ of alpha particles, consistent with enhanced He$^{+}$ often reported in ICMEs \citep{2003JGRA..108.8040K,2025ApJ...981...35O}. Because the density is exceptionally low, the SWA-PAS moments show large variability likely caused by counting statistics and instrumental limitations. The PAS quality factor is $>$ 0 (untrustworthy) for over 25\% of time steps during the SEP interval \citep[see also][]{2022JGRA..12730754D}. We therefore do not attempt to relate short-timescale plasma variability to IMF fluctuations during this period and simply note that the densities remain low throughout the ICME passage, which ends near 12:00\,UT on December 25.

Figure~\ref{fig:mu}h shows the IMF magnitude and components. $B_\mathrm{R} > 0$ throughout the ICME, while $B_\mathrm{N}$ decreases gradually from $\sim$5\,nT to $\sim-5$\,nT. Beginning around 10:15\,UT on December 24, the IMF becomes strongly radial for an extended interval that overlaps a sequence of later dropouts. Quantifying the radiality using $B_\mathrm{R}/|B|$ at 16\,s cadence from 10:15\,UT to 22:00\,UT yields an average radial fraction of 96.3\%$\pm$1.2\%. 

Finally, Fig.~\ref{fig:mu}i shows the 91\,eV suprathermal electron PADs from SWA-EAS. Prior to the ICME arrival, the electrons are largely unidirectional away from the Sun along the IMF. At the ICME onset a counter-streaming population becomes apparent. Electron intensities then decrease gradually through the early SEP interval and become strongly depleted during the first SEP dropout, most clearly at 50--100\,eV (the depletion is less pronounced at $>100$\,eV; not shown). During the later dropout sequence in the near-radial IMF interval, the suprathermal electron intensities show intervals of depletion (marked by the dashed black lines in Fig.~\ref{fig:mu})  that tend to alternate with SEP ion dropouts.

\section{Discussion and conclusions}\label{sec:dis}

Using Parker-spiral ballistic back-mapping to $r_\mathrm{SS} = 2.5$\,R$_{\odot}$ together with PFSS extrapolations, we find that Solar Orbiter’s nominal photospheric footpoint and SO$-$10 remain within the jet-base/source region from the inferred ion release ($\sim$04\,UT) through the main dropout interval ($\sim$07:15--10\,UT). In contrast, the SO$+$10 is markedly more sensitive and frequently connects to open field lines outside the jet base whose source surface endpoints lie within the ecliptic band. In the QSL-proxy maps this behavior is associated with nearby strong connectivity gradients, indicating that small displacements in the inferred footpoint can correspond to connectivity changes between neighboring open-field regions of the same polarity. This topology provides a plausible basis for interpreting brief dropouts as transitions between adjacent open flux tubes with different access to particles released from the compact coronal jet/source region, but quasi-static global PFSS extrapolations cannot determine whether such switching occurred in this event. Resolving the dropout mechanism more directly will require time-dependent coronal modeling and/or independent observational constraints on how the open-flux connectivity evolved near the source.

The main SEP dropout between 07:15 and 10:00\,UT is not accompanied by an abrupt change in local IMF direction or in reliable solar wind moments at the dropout onset. The sequence of dropouts in this event is consistent with earlier statistics indicating that many dropouts are not associated with large IMF changes \citep[e.g.,][]{2008ApJ...688.1368C}. Among the dropout edges identified in Fig.~\ref{fig:mu}, only a minority coincide with abrupt in-situ changes, such as the transition into the strongly radial IMF interval near 10:15\,UT and a late event edge near 23:45\,UT associated with a solar wind speed and density jump. 

The gradual, energy-dependent decrease of the suprathermal electrons suggests reduced access to the suprathermal-electron population and/or sampling of flux tubes with different connectivity. The electron data alone do not uniquely support a fully disconnected or closed-loop interpretation. The alternating suprathermal electron and SEP ion depletion patterns during the later dropout sequence suggest patchy connectivity, in which Solar Orbiter repeatedly samples adjacent interleaved flux tubes \citep{2014ApJ...780...16G} with different particle access. 

The immediate disappearance of the field-aligned SEP beam together with the sudden appearance of a weak 100--200\,keV EPT ion population over $\alpha\sim$90$^\circ$--180$^\circ$ is consistent with Solar Orbiter sampling adjacent flux tubes with different connectivity and SEP access, such that a weak low-energy component can persist while the MeV ions show no similar component over this pitch-angle range. This weak component could reflect a suprathermal population already present on adjacent flux tubes and/or a weak event-related population.

Finally, the dropout-rich interval occurs during a magnetic-cloud passage with reduced magnetic fluctuations. Such conditions may limit cross-field transport, helping maintain sharp SEP intensity gradients between neighboring flux tubes and making dropouts more apparent. We therefore interpret the magnetic cloud primarily as providing favorable conditions for the dropout. This single-event case study cannot determine whether magnetic-cloud intervals commonly accompany SEP dropouts. Addressing this will require a larger sample of dropout events observed under different solar wind conditions. This interpretation is qualitatively consistent with \citet{2023A&A...678A..98W}, who reported modest flux dropouts in a Solar Orbiter event inside an ICME/flux-rope interval and suggested that low IMF fluctuations helped preserve a narrow anisotropic beam.

\begin{acknowledgments}
R.B. \rev{and M.A.D.} acknowledge support by NASA grants 80NSSC22K0757, 80NSSC24K1441 \rev{and 80NSSC25K7687}. \rev{R.G.H. acknowledges the financial support by project PID2023-150952OB-I00 funded by MICIU/AEI/10.13039/501100011033 and EU/FEDER and project SBPLY/24/180225/000108 funded by EU/FEDER and JCCM/INNOCAM.}
\end{acknowledgments}

\software{SolarSoft \citep{1998SoPh..182..497F}
          }


\bibliography{ads}{}
\bibliographystyle{aasjournalv7}



\end{document}